\documentclass[12pt,draftcls,onecolumn]{IEEEtran_new}
\usepackage{color}
\usepackage{amsmath,amssymb,amscd}
\usepackage{subfigure, epsfig}

\newtheorem{theorem}{Theorem}
\newtheorem{definition}[theorem]{Definition}
\newtheorem{proposition}[theorem]{Proposition}

\newcommand{\mb}{\mathbf}
\newcommand{\ul}{\underline}

\newcommand{\R}{\mathbb{R}}
\newcommand{\PP}{\mathcal{P}}

\newcommand{\mc}{\mathcal}

\newcommand{\ds}{\displaystyle}

\ifCLASSINFOpdf
\else
\fi
\begin{document}

\title{A Repeated Game Formulation of
Energy-Efficient Decentralized Power Control}

\author{M.~Le~Treust and S.~Lasaulce}

\maketitle

\begin{abstract}
Decentralized multiple access channels where each transmitter wants
to selfishly maximize his transmission energy-efficiency are
considered. Transmitters are assumed to choose freely their power
control policy and interact (through multiuser interference) several times. It is shown that the
corresponding conflict of interest can have a predictable outcome,
namely a finitely or discounted repeated game equilibrium. Remarkably,
it is shown that
 this equilibrium is Pareto-efficient under reasonable sufficient conditions
 and the corresponding decentralized power control policies can be implemented under
  realistic information assumptions: only individual channel state
 information and a public signal are required to implement the equilibrium strategies. Explicit
 equilibrium conditions are derived in terms of minimum number of game stages or maximum
 discount factor. Both analytical
 and simulation results are provided to compare the performance of the proposed
 power control policies with those already existing and exploiting the same
 information assumptions namely, those derived for the one-shot
 and Stackelberg games.
\end{abstract}


\section{Introduction}
\label{sec:intro}

Many current wireless communications systems (e.g., cellular
networks) are optimized in terms of quality of service (QoS), which
can include for example, performance criteria such as transmission
rate, reliability, latency, or security. It turns out that
applications where trade-offs have to be found between QoS and
energy consumptions, have become more and more important, especially
over the past decade. Wireless sensor networks and ad-hoc networks
are two good examples illustrating the importance of finding such
trade-offs. A very simple and pragmatic way of knowing to what
extent a communication is energy-efficient has been proposed by
\cite{shah-pimrc-1998}\cite{goodman-pc-2000}. The authors of
\cite{shah-pimrc-1998}\cite{goodman-pc-2000} define
energy-efficiency as the net number of information bits that are
transmitted without error per unit time (goodput) to the transmit
power level. More specifically, the authors analyze the problem of
distributed power control (PC) in flat fading multiple access
channels (MACs). The problem is formulated as a non-cooperative game
where the players are the transmitters, the action of a given player
is her/his/its transmit power (``his'' is chosen in this paper), and
his payoff/reward/utility function is the energy-efficiency of his
communication with the receiver. The results reported in
\cite{shah-pimrc-1998}\cite{goodman-pc-2000} have been extended to
the case of multi-carrier systems in \cite{meshkati-jsac-2006}.
Unfortunately, as shown in \cite{goodman-pc-2000}, Nash equilibria
(NE) resulting from the one-shot game formulation of the
energy-efficient power control problem are generally inefficient.
This is one of the reasons why some authors proposed to apply other
game-theoretic concepts to improve efficiency of the network
equilibrium: \cite{saraydar-com-2002} proposed a pricing mechanism
and \cite{lasaulce-twc-2009} proposed to introduce some hierarchy in
the network by considering a Stackelberg game
\cite{stackelberg-book-1934} or using successive interference
cancellation at the receiver. The solution of
\cite{saraydar-com-2002} has the advantage to be Pareto-optimal (PO)
but requires global channel state information (CSI) at the
transmitters and equilibrium uniqueness is not proven analytically.
On the other hand, the solution of \cite{lasaulce-twc-2009} is not
PO but only requires individual CSI and uniqueness is guaranteed.

All the cited and related works on energy-efficient PC
(\cite{meshkati-jsac-2006}\cite{saraydar-com-2002}\cite{lasaulce-twc-2009}, etc) have at least one common point: time
is divided into windows or blocks over which the channel is assumed to be
constant and transmit power levels can be updated only once within a
given block. The corresponding framework is the one of static
or one-shot games which is to say,
transmitters play independently from block to block and maximize
their instantaneous utility for each block. In this paper, we consider a more
general situation: transmitters are allowed to update their
power levels several times within a block; the corresponding
 PC type could be called
decentralized fast PC (DFPC), generalizing the more conventional
decentralized slow PC (DSPC) for which the power can be updated only
once per block. Both in the DFPC and DSPC cases, we want to take
into account the fact that players (namely the transmitters)
interact several times within a block or/and from block to block,
 which introduces new types of behaviors (cooperation, punishment, etc)
with respect to the one-shot game. The framework considered here is
the one of dynamic games. More specifically, we analyze a special
case of dynamic games, which is the case of repeated games (RG). In
standard repeated games
\cite{aumann-book-1981}\cite{sorin-handbook-1992}\cite{msz} the same
game is played a finite or infinite number of times and players are
interested in optimizing a certain performance metric, resulting
from averaging their utility over the whole duration of the game. In
contrast with iterative or learning techniques that
     are based on mild information assumptions and different
     behavior assumptions (transmitters can be modeled by
     automata), RG generally require more demanding information and
     behavior assumptions. Also, RG aim at optimizing an averaged
 utility and reaching points more efficient than the one-shot game NE. Two important models of RG are considered: the finitely
repeated game (FRG) \cite{BenoitKrishna85} and discounted repeated
game (DRG) \cite{fud}. A priori, the FRG seems to be more suited to
DFPC since the number of times players interact is finite and can be
known (e.g., the number of training symbols in a block) whereas the
DRG seems to be more suited to DSPC with a uncertain number of
blocks over which players interact. To the authors' knowledge, there
is only a small fraction of papers dedicated to repeated games in
the wireless literature. As far as the present paper is concerned,
the most relevant contributions available are
\cite{etkin-jsac-2007}\cite{lai-it-2008}\cite{wu-twc-2009}. With
respect to these works, our contributions are as follows. The work
reported in this paper is the first to apply the concept of RG to
energy-efficient PC (the existing works consider Shannon
transmission rates or similar utility functions). A second important
feature of the present analysis is that only individual CSI and
signal-to-interference plus noise ratio (SINR) are needed to
implement the proposed PC scheme, which is not the case in
\cite{etkin-jsac-2007}\cite{lai-it-2008}\cite{wu-twc-2009}. Third,
two models of RG are considered and explicit equilibrium conditions
on the number of game stages (for the FRG) and discounted factor
(for the DRG) are provided and discussed, which is not made in the
existing
 literature. At last but not least, the PC policy we propose is compared in a fair manner
 with existing game-theoretic PC policies both analytically and by simulations
 and shown to be the most efficient one (in the sense of Pareto).
 For this purpose, several works
 from the game theory literature, and not used yet by the wireless community, are exploited.

This paper is structured as follows. In Sec. \ref{sec:signal-model}
the assumed signal model is described. This is followed (Sec.
\ref{sec:game-definition}) by a short review of the static/one-shot
non-cooperative and Stackelberg PC games. In Sec. \ref{sec:repeated-game} a
rigorous RG formulation of the PC problem is provided and
information assumptions necessary to implement the equilibrium
strategies of Sec. \ref{sec:eq-strategies} are given. The proposed equilibria (finitely and discounted
repeated games equilibria) and their properties are
analyzed in Sec. \ref{sec:equilibrium-rg}. The results derived are
illustrated by simulations in Sec. \ref{sec:numerical-results},
which is followed by the conclusion (Sec. \ref{sec:conclusion}).


\section{System model}
\label{sec:system-model}

\subsection{Signal model} \label{sec:signal-model}

We consider a decentralized MAC with a finite number of
transmitters, which is denoted by $K$. The network is said to be
decentralized in the sense that the receiver (e.g., a base station)
does not dictate to the transmitters (e.g., mobile stations) their
PC policy. Rather, all the transmitters choose their policy by
themselves and want to selfishly maximize their energy-efficiency;
in particular, they can ignore some specified centralized policies.
We assume that the users transmit their data over quasi-static
channels and at the same time and frequency band. Note that a block
is defined as a sequence of $M$ consecutive symbols which comprises
a training sequence that is, a certain number of consecutive symbols
used to estimate the channel (or other related quantities)
associated with a given block. A block has therefore a duration less
than the channel coherence time. The signal model used corresponds
to the information-theoretic channel model used for studying MAC
\cite{wyner-it-1974}\cite{cover-book-1975}; see e.g.,
\cite{belmega-twc-2009}
 for more comments on the multiple access
technique involved. What matters is that this model is both simple
to be presented
 and captures the different aspects of the
problem (the SINR structure in particular) and can be readily
applied to speficic systems such as CDMA systems \cite{goodman-pc-2000}\cite{lasaulce-twc-2009}
 or multi-carrier CDMA systems \cite{meshkati-jsac-2006}. The equivalent baseband signal
received by the base station can be written as
\begin{equation}
\label{eq:received-signal}
 y(n) = \sum_{i=1}^{K} g_i(n) x_i(n) + z(n)
\end{equation}
where $i  \in \mc{K}$, $\mc{K} = \{1,...,K\}$, $x_i(n)$ represents
the symbol transmitted by transmitter $i$ at time $n$,
$\mathbb{E}|x_i|^2 = p_i$, the noise $z$ is assumed to be
distributed according to a zero-mean Gaussian random variable with
variance $\sigma^2$ and each channel gain $g_i$ varies over time but
is assumed to be constant over each block. For each transmitter $i$,
the channel gain modulus is assumed to lie in a compact set
$\left|g_i \right|^2 \in \left[\eta_i^{\mathrm{min}},
\eta_i^{\mathrm{max}} \right]$. This assumption is both practical
(for example, such limitations can model the finite receiver
sensitivity and the existence of a minimum distance between the
transmitter and receiver) and important to guarantee the existence
of the proposed equilibria (Sec. \ref{sec:equilibrium-rg}). At last,
the receiver is assumed to implement single-user decoding.

\subsection{Review of the one-shot power control game}
\label{sec:game-definition}

Here, we review a few key results from \cite{goodman-pc-2000}
concerning the static non-cooperative PC game. In order to define
the static PC game some notations need to be introduced. We denote
by $R_i$ the transmission information rate (in bps) for user $i$ and
$f$ an efficiency function representing the block success rate,
which is assumed to be sigmoidal and identical for all the users;
the sigmoidness assumption is a reasonable assumption, which is well
justified in
\cite{rodriguez-globecom-2003}\cite{meshkati-tcom-2005}. Recently,
\cite{belmega-valuetools-2009} has shown that this assumption is
also justified from an information-theoretic standpoint. At a given
instant,
 the SINR at receiver $i \in \mc{K}$ writes as:
\begin{equation}
\label{eq:sinr-ne} \mathrm{SINR}_i=\frac{p_i|g_i|^2}{\sum_{j \neq i}
p_j|g_j|^2+\sigma^2}
\end{equation}
where $p_i$ is the power level for transmitter $i$. With these
notations, the static PC game, called $\mc{G}$, is defined in its
normal form as follows.

\begin{definition}[Static PC game] \emph{The static PC game is a triplet
$\mc{G} = (\mc{K}, \{\mc{A}_i\}_{i\in\mc{K}},\{u_i\}_{i\in\mc{K}})$
where $\mc{K}$ is the set of players, $\mc{A}_1,...,\mc{A}_K$ are
the corresponding sets of actions, $\mc{A}_i = [0,
P_i^{\mathrm{max}}]$, $P_i^{\mathrm{max}}$ is the maximum transmit
power for player $i$, and $u_1,...,u_k$ are the utilities of the
different players which are defined by:}
\begin{eqnarray}
\label{eq:def-of-utility} u_i(p_1,...,p_K)= \frac{R_i
f(\mathrm{SINR}_i)}{p_i} \ [\mathrm{bit} / \mathrm{J}].
\end{eqnarray}
\end{definition}
In this game with complete information ($\mc{G}$ is known to every
player) and rational players (every player does the best for himself
and knows the others do so and so on), an important game solution
concept is the NE (i.e., a point from which no player has interest
in unilaterally deviating). When it exists, the non-saturated NE of
this game can by obtained by setting $\frac{\partial u_i}{\partial
p_i}$ to zero, which gives an equivalent condition on the SINR: the
best SINR in terms of energy-efficiency for transmitter $i$ has to
be a solution of $xf'(x)-f(x)=0$ (this solution is independent of
the player index since a common efficiency function is assumed, see
\cite{meshkati-tcom-2005} for more details). This leads to:
\begin{equation} \forall i \in \{1,...,K \}, \
p_i^{*}= \frac{\sigma^2}{|g_i|^2} \frac{\beta^*}{1-(K-1)\beta^{*}}
\label{eq:NE-power}
\end{equation}
where $\beta^{*}$ is the unique solution of the equation
$xf'(x)-f(x)=0$. By using the term ``non-saturated NE'' we mean that
the maximum transmit power for each user, denoted by $
P_i^{\mathrm{max}}$, is assumed to be sufficiently high not to be
reached at the equilibrium i.e., each user maximizes his
energy-efficiency for a value less than $P_i^{\mathrm{max}}$ (see
\cite{lasaulce-twc-2009} for more details). An important property of
the NE given by (\ref{eq:NE-power}) is that transmitters only need
to know their individual channel gain $|g_i|$ to play their
equilibrium strategy. One of the interesting results of this paper
is that it is possible to obtain a more efficient equilibrium point
by repeating the game $\mc{G}$ while keeping this key property.

\subsection{Review of the Stackelberg power control game}
\label{sec:review-stackelberg}

Here we review a few key results from \cite{lasaulce-twc-2009}. The
corresponding results will be used in Sec. IV and V where the
performance of the proposed RG equilibrium is compared with the one
of the Stackelberg equilibrium (SE). The main difference between the
one-shot game \cite{goodman-pc-2000} and the Stackelberg game is
that in the latter, one of the transmitters is the leader of the
game. The other transmitters are assumed to be able to observe the
actions of the leader and react to them accordingly; these
transmitters are called the followers. Note that the leader knows
that his actions are observed and therefore anticipates the reaction
of the followers. In \cite{lasaulce-twc-2009}, it is shown that the
SE power profile Pareto-dominates the one-shot NE power profile. The
power played by transmitter $i \in \mc{K}$ at the SE when he is the
game leader (L) is given by: $p_i^L=\frac{\sigma^2}{|g_i|^2}
\frac{\gamma^*(1+\beta^*)}{1-(K-1)\gamma^*\beta^*-(K-2)\beta^*}$
where $\gamma^*$ is the unique solution of $x
\left[1-\frac{(K-1)\beta^*}{1-(K-2)\beta^*} x\right] f'(x) - f(x)
=0$, and the corresponding utility is
$u^L_i=\frac{|g_i|^2}{\sigma^2}\frac{1-(K-1)\gamma^*\beta^*
-(K-2)\beta^*}{\gamma^*(1+\beta^*)}f(\gamma^*)$. On the other hand,
the power played by transmitter $i \in \mc{K}$ at the SE when he is
one of the game followers (F) and corresponding utility are given by
$p_i^F=\frac{\sigma^2}{|g_i|^2}\frac{\beta^*(1
+\gamma^*)}{1-(K-1)\gamma^*\beta^*-(K-2)\beta^*}$ and
$u^F_i=\frac{|g_i|^2}{\sigma^2}\frac{1-(K-1)\gamma^*\beta^*
-(K-2)\beta^*}{\beta^*(1+\gamma^*)}f(\beta^*)$.

\section{Moving from static to repeated PC games}
\label{sec:repeated-game}

\subsection{Strategic information assumptions}
\label{sec:RG-info-assumptions}


First, we want to show that, in the non-saturated regime (as defined
in the preceding section), the energy-efficient one-shot PC game has
a very interesting structure in terms of CSI. This structure, which
is analyzed just below, has two consequences. The first consequence
is that
 only individual CSI is required at the transmitters that is,
only $|g_i|$ needs to be known by transmitter $i$. In the existing
literature (\cite{goodman-pc-2000}\cite{meshkati-jsac-2006}\cite{lasaulce-twc-2009},
 etc) this game feature is observed once
the equilibrium solution has been determined. It turns out that it
is due to the structure of the utility functions of the one-shot
game and not only to the specific solutions analyzed in the existing
literature. This can therefore be checked before deriving the
equilibrium solution. A simple way of proving this statement is to
consider a game where utilities are normalized:
$\frac{u_i(p_1,...,p_K)}{|g_i|^2}=\frac{R_i
f\left(\frac{p_i|g_i|^2}{\sum_{j\not=i}p_j|g_j|^2+\sigma^2}\right)}{p_i|g_i|^2}$.
By
 making the change of variables $a_i=p_i|g_i|^2 $ the normalized
utility function becomes: $\hat{u}_i(a_1,...,a_K)= \frac{R_i
f\left(\frac{a_i}{\sum_{j\not=i}a_j+\sigma^2}\right)}{a_i}$. It is seen from the normalized utilities that channel gains play only a role when players de-normalize their actions
 by computing $p_i = \frac{a_i}{|g_i|^2}$, which shows that only individual CSI is needed to
  play the game. The second consequence of the structure of
  $\hat{u}_i$ is that the repeated versions of the one-shot game are
   easier to be analyzed. Indeed, the normalized utilities $\hat{u}_i$ depend
   on time only through the action profile $\ul{a} = (a_1, ...,a_K)$ and not
   through $|g_i|$. This means that the PC stochastic repeated game
    can be analyzed as the repeated version of a (normalized) static game; this is why,
    in the sequel, we will not use a time index  for $u_i$. In addition to the individual CSI
   assumption, we also assume that every player can observe a public
signal. More specifically, we assume that every player knows at each
stage of the game the signal
\begin{equation}
\label{eq:def-public-signal}
\omega = \sigma^2 + \sum_{ i=1 }^K |g_i | ^ 2  p_i.
\end{equation}
By noticing that $\forall i \in \mc{K}, \  p_i |g_i|^2 \times \frac{\mathrm{SINR}_i +1}{\mathrm{SINR}_i}
= \omega$ we see that each transmitter can construct the public signal from the sole knowledge of his action,
individual channel gain, and individual SINR. One of the main results of this paper is that assuming the two mentioned
 information assumptions allows one to implement a cooperation plan between
 the transmitters corresponding to an efficient equilibrium. This is
 possible with the repeated game formulation of the problem, which
 is given below.

\subsection{Repeated game formulation of the power control problem}
\label{sec:RG-formulation}

In the static PC game, each transmitter observes the channel gain
associated with the current block i.e., $|g_i|$ and updates his
power level according to (\ref{eq:NE-power}) in order to maximize
his instantaneous utility. In repeated games, players want to
maximize their averaged utility. To define the latter quantity we
first need to define several notions namely a game stage, the game
history, and the strategy of a player. Game stages correspond to
instants at which players can choose their actions. To be concrete,
in the case of DSPC, game stages coincide with blocks whereas in the
case of DFPC, game stages coincide with sub-blocks (comprising one
or several symbols). The proposed framework is the one of repeated
games with a public signal (see e.g., \cite{Fudenberg-Levine-Maskin1994}\cite{tomala-ijgt-98}) in
which the public signal (namely (\ref{eq:def-public-signal})) is a
deterministic function of the played actions. We denote by $\Omega
=\ds{ \left[\sigma^2, \sigma^2 + \sum_{i=1}^{K} \eta_i^{\max}
P_i^{\mathrm{max}} \right]}$ the interval the public signal lies in. The
pair of vectors $(\ul{\omega}_t, \ul{p}_{i,t}) =
\left(\omega(1),...,\omega(t-1), p_{i}(1),...,p_i(t-1)\right)$ is
called the history of the game for transmitter $i$ at time $t$ and
lies in the set $\mc{H}_t = \Omega^{t-1} \times \mc{P}_i^{t-1}$.
This is precisely the history $\ul{h}_t = \left(\ul{\omega}_t,
 \ul{p}_{i,t}\right)$ which is assumed to be known by the
  transmitters before playing at stage $t$. With
these notations, a pure strategy of a transmitter in the repeated
game can be defined properly.
\begin{definition}[Players' strategies in the RG] \emph{A pure strategy $\tau_i$ for
player $i \in \mc{K}$ is a sequence of causal functions
$\left(\tau_{i,t} \right)_{t \geq 1}$ with}
\begin{equation}
\label{eq:strategy-rg} \tau_{i,t}: \left|
\begin{array}{ccc}
\mc{H}_t & \rightarrow & [0, P_i^{\mathrm{max}}]\\
 \ul{h}_t & \mapsto & p_i(t).
\end{array}
\right.
\end{equation}
\end{definition}
The strategy of player $i$, which is a sequence of functions, will
be denoted by $\tau_i$ (removing the game stage index in
(\ref{eq:strategy-rg}) is a common way to refer to a player's
strategy). The vector of strategies $\ul{\tau} = (\tau_1, ...,
\tau_K)$ will be referred to a joint strategy and lie in the set $\mc{T}$. A joint strategy
$\ul{\tau}$ induces in a natural way a unique action plan
$(\ul{p}(t))_{t\geq 1}$. To each profile of powers $\ul{p}(t) =
(p_1(t),\hdots,p_K(t))$ corresponds a certain instantaneous utility
$u_i(\ul{p}(t)) $ for player $i$. In our setup, each player does not
only care about what he gets at a given stage but what he gets over
the whole duration of the game. This is why we consider utility
functions resulting from averaging over the instantaneous utility.
More precisely, we consider two utility functions which correspond
to the two important models of repeated games under investigation
namely the FRG and DRG.

\begin{definition}[Players' utilities in the RG] \emph{Let $\ul{\tau} = (\tau_1, ..., \tau_K)$
be a joint strategy. The utility for player $i \in \mc{K}$ is
defined by:}
\begin{equation}
\label{eq:utility-rg}
\begin{array}{ccll}
v_i^T(\ul{\tau}) & = & \ds{\frac{1}{T}\sum_{t =1}^T
 u_i(\ul{p}(t))} & \ \text{\emph{in the FRG}} \\
 v_i^{\lambda}(\ul{\tau}) & = & \ds{\sum_{t =1}^{\infty}
\lambda (1 - \lambda)^{t-1} u_i(\ul{p}(t))} & \ \text{\emph{in the DRG}}
\end{array}
\end{equation}
\emph{where $\ul{p}(t)$ is the power profile of the action plan
induced by the joint strategy $\ul{\tau}$, $T\geq1$ is the number of
game stages in the FRG, and $0<\lambda<1$ is a parameter of the DRG
called the discount factor and is known to every player (since the game is
with complete information).}
\end{definition}
In this paper, we consider two models of RG because these are
complementary in a certain sense. In scenarios where the duration
over which the transmitters interact is known (this is typically the
case when transmit power levels can only be updated during the
training phase of the block) or/and transmitters value the different
instantaneous
 utilities uniformly, the FRG seems to be more suited. In the current available wireless
  literature on the problem under
investigation the DRG is used as follows: the discount factor is
used in \cite{etkin-jsac-2007} as a way of accounting for the delay
sensitivity of the network; the discount factor is used in
\cite{wu-twc-2009} to let the transmitters the possibility to value
short-term and long-term gains differently. Interestingly,
\cite{sorin-handbook-1992}\cite{shapley-pnas-1953} offer another
interpretation of this model. Indeed, the author sees the DRG as an
FRG where $T$ would be unknown to the players and considered as an
integer-valued random variable, finite almost surely, whose law is
known by the players. Otherwise said, $\lambda$ can be seen as the
stopping probability at each game stage: the probability that the
game stops at stage $t$ is thus $\lambda (1-\lambda )^{ t-1 }$. The
function $v_i^{\lambda}$ would correspond to an expected utility
given the law of $T$. This shows
 that the discount factor is also useful to study wireless
games where a player enters/leaves the game. We
 would also like to mention that it can also model a heterogeneous DRG where
 players have different discount factors, in which case $\lambda$ represents
 $\min_i \lambda_i$ (as pointed out in \cite{shapley-pnas-1953}). In
 practice, such a parameter can be acquired through a public signal from the receiver.

\begin{definition}[Equilibrium strategies in the RG] \emph{A joint strategy $\ul{\tau}$ supports
 an equilibrium of the RG defined by $\left(\mc{K},
  \{\mc{T}_i\}_{i \in \mc{K}}, \{v_i\}_{i \in \mc{K}} \right)$
if $\forall i \in \mc{K}, \forall \tau_i' \in \mc{T}_i, \
v_i(\ul{\tau}) \geq v_i(\tau_i', \ul{\tau}_{-i})$ where $v_i$ equals
$v_i^T$ or $v_i^{\lambda}$, $-i$ is the standard notation to refer
to the set $\mc{K} \backslash \{i\}$; here $\ul{\tau}_{-i} =
(\tau_1, ..., \tau_{i-1}, \tau_{i+1},...,\tau_K)$.}
\end{definition}
An important issue is precisely to characterize the set of possible
equilibrium utilities in the RG. When the game is with complete
information and perfect monitoring, there is a theorem which
provides the corresponding characterization from the sole knowledge
of the possible utilities in the static game. This theorem is called
the ``Folk theorem'' (see e.g.,
\cite{aumann-book-1981}\cite{sorin-handbook-1992}). Recall that the
game is said to be with perfect monitoring when every player is able
to observe at each stage the actions chosen by all the other
players. Whereas this knowledge can be acquired in certain scenarios
where appropriate estimation and sensing mechanisms are implemented,
one of our objectives is to show that some of these information
assumptions can be relaxed by exploiting the specific structure of
energy-efficient PC games. In the next section, we clearly make explicit
the information assumptions needed to implement the proposed
cooperation plan. This cooperation plan is built from a point of the
feasible utility region of $\mc{G}$ and has been studied in
\cite{goodman-mna-2001} in the framework of centralized networks.

\section{Equilibrium analysis of the repeated PC game}
\label{sec:equilibrium-rg}

Folk theorems aim at
characterizing the set of equilibrium utilities of RG for different types of RG
 (finite/infinite/discounted RG), equilibria (NE, correlated equilibria,
 communication equilibria, etc), and information assumptions
 (complete/incomplete information, perfect/imperfect monitoring). In this paper, the objective is much
 more modest since we focus on a given
 equilibrium point and a specific game. One of the goals of this section
 is to show that the proposed equilibrium has several
 attractive properties for wireless networks: (a) as shown in the previous section, only individual CSI is required at
    the transmitters; (b) it is fair in terms of SINR like the NE in the one-shot
    game $\mc{G}$; (c) it is PO under sufficient but reasonable conditions; (d) it
     is more efficient than the equilibrium point of the SE point of \cite{lasaulce-twc-2009} under sufficient but reasonable conditions; (e)
    it is always more efficient than the NE in the one-shot
    game $\mc{G}$; (f) it is subgame perfect (this notion will be explained
    in Sec. \ref{sec:eq-strategies}) in the case of the DRG. The corresponding equilibrium relies on a cooperation plan exploiting two
points of the one-shot game: the NE point presented in Sec. \ref{sec:game-definition}
and the operating point for which a detailed study is required.

\subsection{An interesting operating point of the
game $\mc{G}$}
\label{sec:point-gamma-tilde}

By considering all the points $(p_1,...,p_K)$ such that $p_i \in [0,
P_i^{\mathrm{max}}]$, $i \in \mc{K}$, one obtains the feasible
utility region. We consider a subset of points of this region for
which the power profiles $(p_1,...,p_K)$ verify
$p_i|g_i|^2=p_j|g_j|^2$ for all $(i,j) \in \mc{K}^2$. The considered
subset therefore consists of the solutions of the following system
of equations:
\begin{equation}
\label{eq:def-operating-point}
\forall (i,j) \in \mc{K}^2, \ \frac{\partial u_i}{\partial p_i}(\ul{p}) = 0
\;\;\text{with }p_i|g_i|^2=p_j|g_j|^2.
\end{equation}
It turns out that, following the lines of the proof of SE uniqueness in \cite{lasaulce-twc-2009}, it is easy to
show that a sufficient condition for ensuring both existence and
uniqueness of the solution to this system of equations is that there
exists $ x_0 \in ]0,\frac{1}{K-1}[$ such
 that $\frac{f''(x)}{f'(x)}-\frac{2(K-1)}{1-(K-1)x}$ is strictly positive on $]0,x_0[$ and strictly negative
 on $]x_0,\frac{1}{K-1}[$. This condition is satisfied for the two following efficiency functions:
$f(x)=(1-e^{-x})^M$ \cite{shah-pimrc-1998} and $f(x) = e^{-\frac{c}{x}}$ \cite{belmega-valuetools-2009}
 with $c = 2^R -1$ ($R$ is the transmission
rate). If this condition should be found to be too restrictive it is always possible to
directly derive a purely numerical condition, which can be translated into a condition on
$K$, $M$ (in the case of \cite{shah-pimrc-1998}), or
$R$ \cite{belmega-valuetools-2009}. Under
the aforementioned condition, the unique solution of
(\ref{eq:def-operating-point}) can be checked to be:
\begin{equation}
\label{eq:power-profile-op-pt}
\forall i \in \mc{K}, \ \tilde{p_i} =
\frac{\sigma^2}{|g_i|^2}\frac{\tilde{\gamma}}{1-(K-1)\tilde{\gamma}}
\end{equation}
where $\tilde{\gamma}$ is the unique solution of $ x[1-(K-1)\cdot x]
f'(x)-f(x)=0$. It is important here to distinguish between the
equal-SINR condition imposed in (\ref{eq:def-operating-point}) and
the equal-SINR solution of the one-shot game (\ref{eq:NE-power}). In
the first case, the SINR has a special structure imposed by the
condition (\ref{eq:def-operating-point}) which is $\mathrm{SINR}_i =
\frac{|g_i|^2 p_i}{\sigma^2 + (K-1)|g_i|^2 p_i}$. Therefore, each
transmitter is assumed to maximize a single-variable utility
function $u_i(\ul{q}_i)$ with $\ul{q}_i = p_i \times
\left(\left|\frac{g_i}{g_1}\right|^2,
\left|\frac{g_i}{g_2}\right|^2, ..., \left|\frac{g_i}{g_K}\right|^2
\right) $. In the second case (one-shot NE), the solution is the
solution of a $K-$unknown $K-$equation system and it happens that
the solution has the equal-SINR property. The proposed operating
point (OP), given by (\ref{eq:power-profile-op-pt}), is thus fair in
the sense of the SINR since $\forall i \in \mc{K}$, $\mathrm{SINR}_i
= \tilde{\gamma}$. Another question to be answered is whether it is
efficient. To answer this question we proceed in three steps.
First, we provide sufficient conditions under which it is PO.
Second, we provide some conditions under which it Pareto-dominates
the SE point derived in \cite{lasaulce-twc-2009}. Third, we show
that the proposed (OP) always Pareto-dominates the one-shot game NE
point (\ref{eq:NE-power}). Denote PO for Pareto-optimality.

\begin{proposition}[Sufficient conditions for PO]\label{prop:PO-sufficient-conditions} \emph{Let
$\mc{U}_{\mc{G}}$ be the achievable utility region for the game $\mc{G}$. Let
$u^{\mathrm{max}}$ be an upper bound for the utilities of all transmitters. Let
$\mc{U}_{\mc{G}}^C$ be the complementary set of $\mc{U}_{\mc{G}}$ in $[0, u^{\mathrm{max}}]^K$. If
$\mc{U}_{\mc{G}}$ or $\mc{U}_{\mc{G}}^C$ is a convex region then the vector of utilities $\ul{u}(\tilde{\ul{p}})$
is Pareto-optimal.}
\end{proposition}
The proof is provided in App. \ref{sec:app-proof-PO}. Whereas it
appears a difficult task to fully characterize the frontier of the
achievable utility region for an arbitrary choice of sigmoidal
functions $f$ and therefore obtain the mildest condition for
Pareto-optimality, many simulations have shown us that with usual
efficiency functions
\cite{shah-pimrc-1998}\cite{belmega-valuetools-2009}, the assumption
``$\mc{G}$ or $\mc{G}^C$ is convex'' is reasonable but
unfortunately not always valid. Of course, the fact
 that we do not provide a general proof does
not mean that the proposed OP is not efficient under milder
conditions. In all the simulations we have performed (not only those
reported in Sec. \ref{sec:numerical-results}), the proposed OP was
PO. The addressed technical problem is in fact a quite general
problem encountered in game theory, especially in economics and some
non-trivial refinements could be brought to our analysis.
\begin{proposition}[SE vs OP]\label{prop:SE-vs-ptilde} \emph{Let $p_i^L$ (resp.
$p_i^F$)
the power of transmitter $i$ at the equilibrium of the Stackelberg game
of \cite{lasaulce-twc-2009} when $i$ is the game leader (resp. one
of the game followers). Denote by $u_i^L$, $u_i^F$ the
corresponding utilities. Then, we have:}
\begin{description}
  \item[\emph{(i)}] $\forall i \in \mc{K}$, \ $u_i(\tilde{\ul{p}}) \geq u_i^L$;
  \item[\emph{(ii)}]$ \exists K_0, \ \forall K \geq K_0, \forall i \in \mc{K},
  u_i(\tilde{\ul{p}}) \geq u_i^F.$
\end{description}
\end{proposition}
The first statement of this proposition indicates that a transmitter
always prefers to play the OP than being a leader of the hierarchical
game of \cite{lasaulce-twc-2009}. The second statement shows that
this is also true for the followers when the network reaches a
certain size in terms of users. All the simulations we have
performed have shown that $K_0$ equals $3$.
\begin{proposition}[NE vs OP]\label{prop:NE-vs-ptilde} \emph{It is always true that $\forall i \in \mc{K},  u_i(\tilde{\ul{p}})
\geq u_i(\ul{p}^*)$ where $\ul{p}^* = (p_1^*,...,p_K^*)$.}
\end{proposition}
An important message conveyed by this proposition is that every
transmitter always prefers the optimal strategy of the RG than the
optimal strategy of the one-shot game. In the next section, we will see that the
proposed OP corresponds to a possible agreement between the players,
which allows one to obtain an equilibrium point in the FRG and DRG.
Note that other points of $\mc{G}$ which are both feasible and
individually rational could be used to build a cooperation plan.
Designing other cooperation plans based on the repeated game
formulation is a relevant extension of this paper.

\subsection{Equilibrium strategies of the repeated PC games}
\label{sec:eq-strategies}
First, we state two theorems providing equilibrium strategies that
have the properties mentioned in Sec. \ref{sec:point-gamma-tilde}. These theorems correspond to the FRG and DRG respectively.
\begin{theorem}[Equilibrium strategies in the FRG]\label{theo:eq-strategies-fini} \emph{Let $T_0$
be some integer. Assume that the following condition is met:} $T \geq T_0$ \emph{with:}
\begin{equation}
\label{eq:cond}
T_0 = \left\lceil \frac{\eta_i^{\mathrm{max}}\frac{f(\beta^*)}{\beta^*} - \eta_i^{\mathrm{min}}
\frac{f(\tilde{\gamma})[1-(K-1)\tilde{\gamma}]}{\tilde{\gamma}}}{\;\eta_i^{\mathrm{min}} \frac{f(\beta^*)[1-(K-1)\beta^*]}{\beta^*}-
 \frac{\eta_i^{\mathrm{max}}f(\beta^*)}{\beta^*\left(\sum_{j \neq i} P_j^{\max}\eta_j^{\mathrm{min}}+\sigma^2\right)}}
  \right\rceil.
\end{equation}
\emph{Then, for all $i \in \mc{K}$, the following action plan is an
NE of the $T-$stage FRG for any distribution for the channel gains
and any $(T, T_0)$ verifying (\ref{eq:cond}) $\forall t \geq 1,$:}
\begin{equation}
\ \tau_{i,t} = \left|
\begin{array}{ll}
\widetilde{p_i} &  \text{\emph{if} } t \in \{1,2,\ldots,T-T_0\} \\
 p_i^* &\text{\emph{if} } t \in \{T-T_0+1, \ldots, T\} \\
P_i^{\max} &\text{ \emph{if someone deviates from }}\\
&\text{ \emph{the above cooperative plan.}}
\end{array}
\right.
 \label{eq:eq-strategies}
\end{equation}
\end{theorem}
\begin{theorem}[Equilibrium strategies in the DRG]\label{theo:eq-strategies} \emph{Assume that
the
following condition is met:}
\begin{equation}
\label{eq:cond2} \lambda \leq
\frac{\eta_i^{\mathrm{min}} \delta(\beta^*, \tilde{\gamma})}{
\eta_i^{\mathrm{min}} \delta(\beta^*, \tilde{\gamma}) + \eta_i^{\mathrm{max}}
 \left[(K-1)f(\beta^*) -  \delta(\beta^*, \tilde{\gamma}) \right]}
\end{equation}
\emph{where $\delta(\beta^*, \tilde{\gamma})
=\frac{1-(K-1)\tilde{\gamma}}{\tilde{\gamma}}
f(\tilde{\gamma})-\frac{1-(K-1)\beta^*}{\beta^*}f(\beta^*)$. Then,
for all $i \in \mc{K}$, the following action plan is a subgame
perfect NE of the DRG for any distribution for the channel gains:}
\begin{equation}
\forall t \geq 1, \ \tau_{i,t} = \left|
\begin{array}{ll}
\widetilde{p_i} & \text{ \emph{if all the other players play} } \tilde{p}_{-i}\\
p_i^* &\text{ \emph{otherwise}}\\
\end{array}.
\right. \label{eq:eq-strategies}
\end{equation}
\end{theorem}
For the proofs see App.
\ref{sec:app-proof-eq-RG}. Here, we restrict our attention to
interpreting these theorems.\\
$\square$ Comment $1$ (Equilibrium conditions). Before starting the
game, the players agree on a certain cooperation/punishment plan.
Each transmitter $i \in \mc{K}$ always transmits at
$\widetilde{p}_i$ if no deviation is detected and plays a punishment
level otherwise ($P_i^{\max}$ or $p_i^*$ depending on the RG under
consideration). The proposed strategies support an equilibrium if
the gain brought by deviating is less than the expected loss induced
by the punishment procedure applied by the other transmitters. To
make sure that this effectively occurs, the game must be
sufficiently long in the FRG and the game stopping probability
sufficiently low in the DRG. This explains
 the presence of the lower bound on $T$ and
upper bound on $\lambda$.\\
$\square$ Comment $2$ (Cooperation plan). In the DRG, the cooperation plan consists in always
transmitting at the powers corresponding to the OP analyzed
in Sec. \ref{sec:point-gamma-tilde}. In the FRG, the cooperation plan includes a phase
where the transmitters play the one-shot game NE, which can seem surprising. The game having a finite number
of stages, it appears that the players who deviate at the last stage
of the game cannot be punished. If a player deviates earlier it can happen
that the punishment undergone by the deviator is not sufficiently
 severe. Therefore, the agreement consisting in playing the one-shot NE during
 the second phase of the cooperation plan corresponds to a selfish trade-off between
 the gain brought by deviating without being punished severely enough and the one
 brought by playing at an efficient point (namely the OP, which Pareto-dominates the one-shot NE).\\
$\square$ Comment $3$ (Deviation detection mechanism). The
cooperation plan is implementable only if the transmitters can
detect a deviation from the cooperation plan. It turns out that the
knowledge of the public signal is sufficient for this purpose.
Indeed, when the transmitters play at the OP, the public signal
equals $\frac{2 \sigma^2}{1 - (K-1) \tilde{\gamma}}$. Therefore, if
one transmitter deviates from the OP, the public signal $\omega(t)$
is no longer constant. Of course, if more transmitters deviate from
the equilibrium in a coordinated manner this detection mechanism can
fail; this is not inherent to the proposed cooperation plan but to
the NE definition. If this issue should turn out to be crucial it
would be necessary to consider other solution concepts such as
strong equilibria which are stable to deviations of coalitions
\cite{Aumann(Core)61}\cite{Mertens80}\cite{sorin-handbook-1992} but
 this is out of scope of this paper.\\
$\square$ Comment $4$ (Punishment procedure). There is one difference between
Theorems $\ref{theo:eq-strategies-fini}$ and $\ref{theo:eq-strategies}$ in
 terms of punishment. In the FRG, the other transmitters punish the
 deviator by playing at their maximum transmit power, which is the most severe punishment
  possible. In the DRG, the punishment is that the other transmitters play at
  the one-shot game NE. The drawback for punishing
at the one-shot game NE point is that the punishment mechanism is
less efficient which is to say, more game stages are needed to punish the deviator. The advantage is that the proposed
equilibrium strategy is subgame perfect \cite {selten1}. 
The subgame perfection property ensures that if a joint strategy $\ul{\tau}$ of
the RG supports an equilibrium then, after all
possible histories $\ul{h}$, the joint strategy $\ul{\tau}(\ul{h})$ is
still an equilibrium strategy, which makes the
 decentralized network performance predictable. Back to the FRG, it can be proven
 \cite{BenoitKrishna85} that the
 subgame perfect equilibrium property cannot be verified in this RG because the
 associated one-shot game has only one pure NE
 (whereas at least two pure NE are needed).\\
$\square$ Comment $5$ (Equilibrium conditions and wireless channels). The equilibrium
conditions provided can be seen to be independent of the channel statistics. If the
 latter are known, it is possible to refine the bounds
  (see App. \ref{sec:app-proof-eq-RG} for more details). To the best of the
 authors' knowledge, this type of conditions are not
 considered in the available wireless literature whereas it is important
 to know whether the models of RG are applicable. The problem of how much
 these conditions are restrictive can be seen
 from two complementary perspectives. If the channel gain
dynamics is given, the question is to know the maximum (resp. minimum)
value of the discount factor (resp. game stages) guaranteeing the existence of an
 equilibrium condition. The values for $\eta_i^{\mathrm{min}}$ and $\eta_i^{\mathrm{max}}$ depend
 on the propagation scenario and considered technology. In systems like WiFi networks
 these quantities typically correspond to the path loss dynamics, the receiver sensitivity, the minimum distance
 between the transmitter and receiver. On the other hand, if the
 discount factor (resp. the number of game stages) is given (e.g., by the
 traffic statistics or training sequence length), this imposes lower and upper bounds on
 the channel gains. It can happen that the admissible
  range for the discount factor (or the number of game stages)
 can be not compatible with the channel gain dynamics, in which case our model
 need to be refined.\\
\section{Numerical results}
\label{sec:numerical-results}
\begin{figure}[h]
\centering
\includegraphics[width=0.80\textwidth]{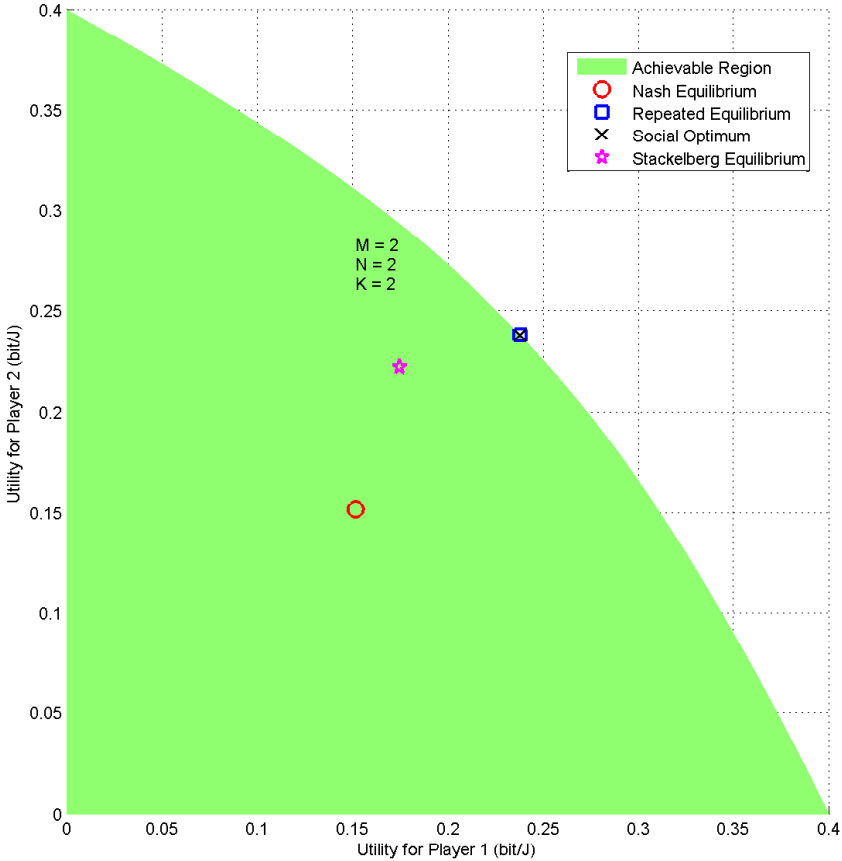}
\caption{Normalized achievable utility region for $(K,M,N)=(2,2,2)$ plus four important
points.}
\label{fig:utility-region-222}
\end{figure}

\begin{figure}[h]
\centering
\includegraphics[width=0.80\textwidth]{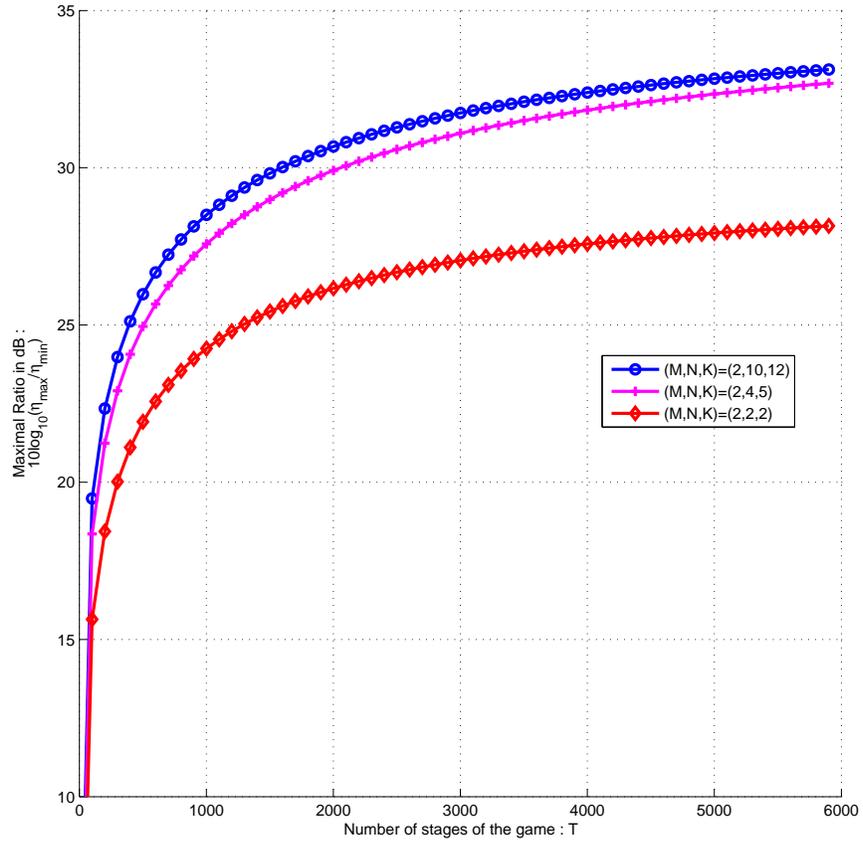}
\caption{Admissible channel gain dynamics vs. number of stages of the game
for $M=2$ and $(K,N) \in \{(2,2),(4,5),(10,12)\}$.}
\label{fig:dynamics-vs-duration}
\end{figure}

\begin{figure}[h]
\centering
\includegraphics[width=0.80\textwidth]{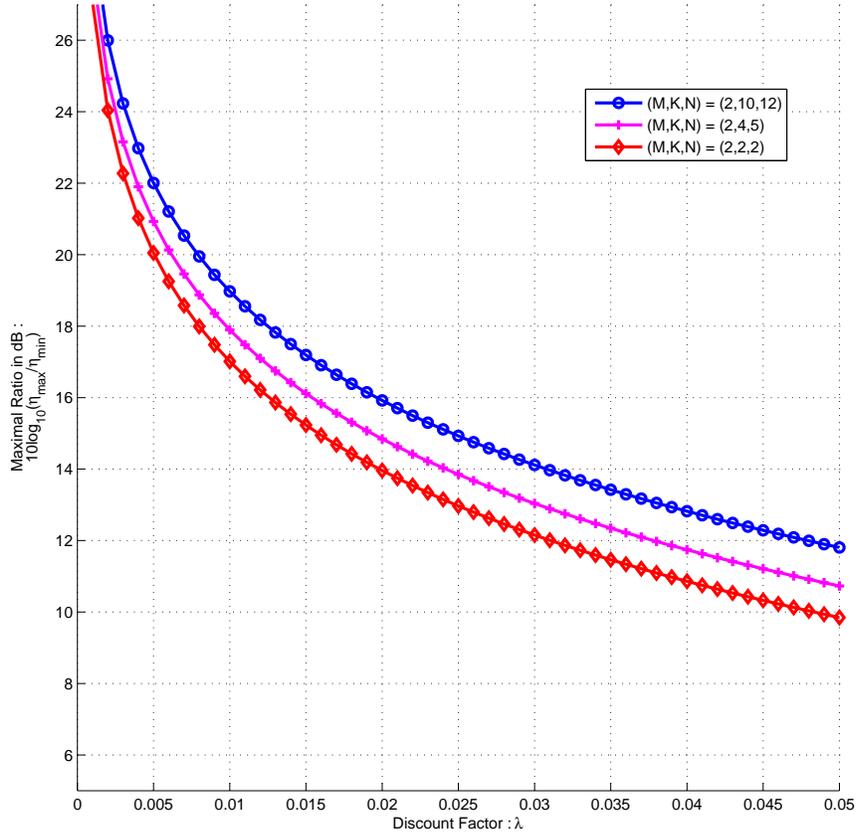}
\caption{Admissible channel gain dynamics vs. discount factor for $M=2$
and $(K,N) \in \{(2,2),(4,5),(10,12)\}$.}
\label{fig:dynamics-vs-lambda}
\end{figure}

\begin{figure}
\centering
\includegraphics[width=0.80\textwidth]{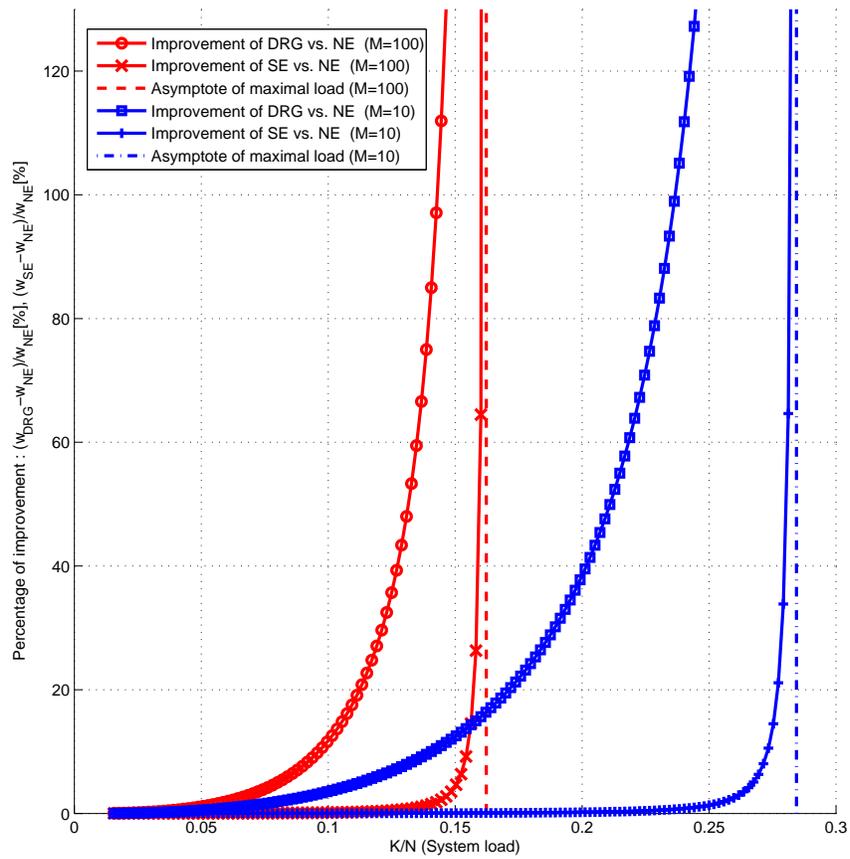}
\caption{Percentage of improvement of social utility for repeated
 equilibrium ($w_{DRG}$) and for Stackelberg ($w_{SE}$) vs. Nash
 equilibrium ($w_{NE}$) in function of the system load ($K/N$).} 
\label{fig:optimality-max-load-disc}
\end{figure}

\begin{figure}
\centering
\includegraphics[width=0.80\textwidth]{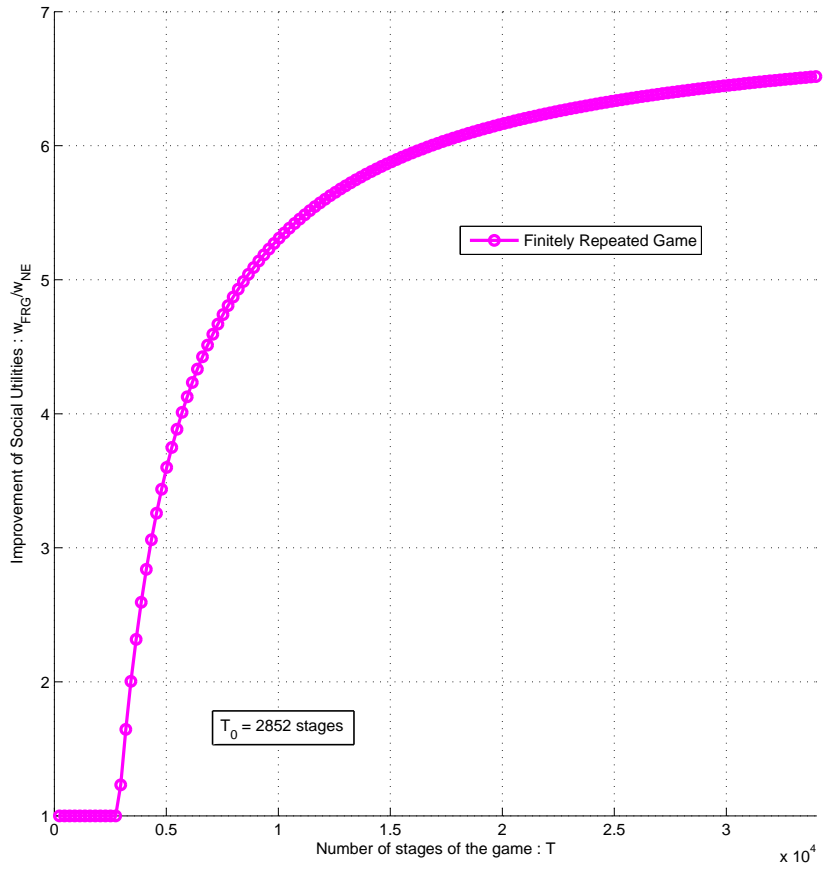}
\caption{Improvement of social utility for the finitely repeated equilibrium ($w_{FRG}$) vs. Nash equilibrium ($w_{NE}$) in function of the number of stages $T$ of the game.}
\label{fig:optimality-max-load-finite}
\end{figure}

In this section we consider the same type of scenarios as
\cite{meshkati-jsac-2006}\cite{lasaulce-twc-2009} namely random CDMA
systems with a spreading factor equal to $N$, the efficiency
function $f(x) = (1- e^{-x})^M$ and Rayleigh fading channels.\\
$\star$ The first scenario considered is simple but has the
advantage that it can be represented clearly. Consider the scenario
$(K,M,N) = (2,2,2)$. Fig. \ref{fig:utility-region-222} represents
the normalized achievable utility region of the one-shot PC game
(normalizing the utilities allows one to conduct fair comparisons).
Four important points are highlighted: the NE of the one-shot game
(circle), the SE (star), the proposed operating/cooperation point
studied in Sec. \ref{sec:point-gamma-tilde} (square marker), and the point
where the social welfare (sum of utilities) is maximized (cross).
From this figure it can be seen that: the utility region is convex;
a significant gain can be obtained by using a model of repeated
games instead of the one-shot model; and the cooperation and optimum
social points coincide.\\
$\star$ Considering the same scenario, the link between the number
of stages of the FRG (resp. the stopping probability of the DRG) and
channel gain dynamics has been considered. Fig.
\ref{fig:dynamics-vs-duration} (resp. \ref{fig:dynamics-vs-lambda})
represents the quantity $10
\log_{10}\left(\frac{\eta^{\mathrm{max}}}{\eta^{\mathrm{min}}}\right)
$ as a function of $T$ (resp. of $\lambda$) for $M=2$ and different numbers of
transmitters and spreading factors : $(K,N) \in
\{(2,2),(4,5),(10,12)\}$. Considering the corresponding figures, the
models of RG seem to be suitable not only in scenarios where $|g_i|$
models the path loss effects but also the fading effects. Of course
if the number of stages is too small or the probability too high,
more appropriate
models have to be designed.\\
$\star$ As a third type of numerical results, the performance gain
brought by the DRG formulation of the distributed PC problem is
assessed. Denote by $w_{NE}$ (resp. $w_{SE}$ and $w_{DRG}$) the
efficiency of the NE (resp. SE and RG equilibrium) in terms of
social welfare i.e., the sum of utilities of the players. Fig.
\ref{fig:optimality-max-load-disc} represents the quantity
$\frac{w_{DRG}-w_{NE}}{w_{NE}}$ and $\frac{w_{SE}-w_{NE}}{w_{NE}} $
in percentage as a function of the spectral efficiency $\alpha =
\frac{K}{N}$ with $N=128$ and $2\leq K<\frac{N}{\beta^*}+1$. The
asymptotes $\alpha_{max} = \frac{1}{\beta^*}+\frac{1}{N}$ are
indicated in dotted lines for different values $M \in \{10,100\}$.
The improvement become very significant when the system load is
close to $\frac{1}{N}+\frac{1}{\beta^*}$, this is because the power
at the
 one-shot game NE becomes large when the system becomes more and more
loaded. As explained in \cite{lasaulce-twc-2009} for the Stackelberg
approach, these gains are in fact limited by the maximum transmit
power.\\
$\star$ At last, Fig. \ref{fig:optimality-max-load-finite}
represents the ratio $\frac{{w}_{FRG}}{w_{NE}}$ as a function of the
number of stages played (averaged over channel gains). Considering the following constants $(K,M,N)=(35,10,128)$, $P^{max}=10^{-2}$ Watt, $\sigma^2=10^{-5}$ Watt, $10\log_{10}(\frac{\eta_{max}}{\eta_{min}})=20$ and the equation (\ref{eq:cond}), a cooperation plan can
be settled as soon as $T_0=2852$ stages. This curve gives an idea of what a transmitter can gain
by cooperating, the normalized gain in terms of utility goes from
$1$ to $6$ depending on the number of stages of the game (between
$2852$ and $15000$ stages). For example, in a cellular system where
power levels are updated with a typical frequency of $1500$ Hz, this
would mean that cooperating is a good option if several transmitters
are using the same resources for more than $1$ s. The ratio
$\frac{{w}_{FRG}}{w_{NE}}$ has a limit when $T\rightarrow +\infty$.
The latter is easy to obtain since
\begin{footnotesize}
\begin{equation}
\frac{{w}_{FRG}}{w_{NE}} = \frac{ \phi(\tilde{\gamma})
\ds{\sum_{i=1}^K R_i} \ds{ \sum_{t=1}^{T-T_0}} |g_i(t)|^2 +
\phi(\beta^*) \ds{\sum_{i=1}^K R_i}\ds{ \sum_{t=T-T_0+1}^{T}
|g_i(t)|^2} }{\phi(\beta^*) \ds{\sum_{i=1}^K R_i} \ds{\sum_{t=1}^T
|g_i(t)|^2}}
\end{equation}
\end{footnotesize}
where $\phi(x) =\frac{f(x)}{x}\left[1-(K-1)x\right]$. It follows
that, for a given $T_0$, $\ds{\lim_{T \rightarrow +
\infty}}\frac{{w}_{FRG}}{w_{NE}} =
\frac{\phi(\tilde{\gamma})}{\phi(\beta^*)}$.

%
%
%
\section{Conclusion}
\label{sec:conclusion}
One of the messages of this paper is that taking into account the
fact that transmitters interact several times, it is possible to incite selfish
transmitters to operate at lower powers, which leads to an
equilibrium point that Pareto-dominates the one-shot NE point of \cite{goodman-pc-2000}
and the SE point of \cite{lasaulce-twc-2009}. It has
been proven that the proposed equilibrium strategies only require
individual CSI and a public signal, which is available in many
wireless systems. Additionally, this equilibrium is also fair in
terms of SINR similarly to the one-shot game NE. In
terms of modeling, two models of RG have been analyzed: the finitely
RG which is suited to situations where the number of game stages
is known, whereas the discounted RG is more suited to situations
where it is uncertain or when typical features of wireless networks
(delay sensitivity of the network, the fact that users
can enter/leave the system, or the fact that transmitters can
 value the current and future utilities differently) have to be
 accounted for. An apparent drawback of these two models
 is the existence of an admissible range for the channel gain dynamics.
 When only path loss is considered, the corresponding effect is generally negligible. If fading
 is also considered, simulations
 have shown that the corresponding impact seems to be
 limited if the game stopping probability (resp. number of game stages) is reasonably low (resp. high). Otherwise,
 the proposed model
 probably needs some refinements such as those
 proposed in \cite{sorin-completeinfo} where the author studies
  the influence of the value of the discount factor on the set
 of possible equilibrium utilities for the prisoners' dilemma. As a more
 general extension of this paper it would
 be important to apply the proposed approach to other network types such as the
 interference channel and studying other equilibrium points (depending on the fairness
 criterion under consideration). At last but not least, the proposed
 communication and game models, even through there are commonly
 used, should be refined to propose cooperation plans more robust to
 imperfect modeling and inherent uncertainty on the quantities used.


\appendices

\section{Proof of Proposition \ref{prop:PO-sufficient-conditions}}
\label{sec:app-proof-PO}

Assume that $\mc{U}_{\mc{G}}$ is convex. Let $\ul{u}$ be a point of
$\mc{U}_{\mc{G}}$. Define the hyperplane $\mc{H}(\ul{u})$,
orthogonal to the vector
$\left(\frac{1}{|g_1|^2},...,\frac{1}{|g_K|^2} \right)$ and
containing $\ul{u}$ by: $\mc{H}(\ul{u}) =\left\{(u_1', ..., u_K')
\in \mathbb{R}_+^{K}: \sum_{i=1}^K \frac{u_i'-u_i}{|g_i|^2}=0
\right\}$. The key observation to be made is that the point
$\ul{\tilde{u}} = (u_1(\tilde{p}),...,u_K(\tilde{p}))$
 given by (\ref{eq:power-profile-op-pt})
 maximizes the weighted sum $\sum_{i=1}^{K} w_i u_i$ if $w_i = \frac{1}{|g_i|^2}$.  Note that $\ul{\tilde{u}}$ is
 unique since the solution of $ x\left[1-(K-1) x \right]f'(x)-f(x)=0$ is
 unique $\tilde{\ul{p}} = \arg \max_{\ul{p}} \sum_{i=1}^{K}
 \frac{u_i(\ul{p})}{|g_i|^2}$. By definition, $\ul{\tilde{u}}$ maximizes the Euclidian distance
$d(\ul{{u}},\ul{0})$ for all points
 on the line originating from $\ul{0}$ and directed by the vector
  $\left(\frac{1}{|g_1|^2},...,\frac{1}{|g_K|^2} \right)$,
  orthogonal to the hyperplane $\mc{H}(\ul{\tilde{u}})$.
In conclusion, $\ul{\tilde{u}}$ maximizes the
         sum $\sum_{i=1}^N\frac{u_i}{|g_i|^2}$, and because of the convexity of the utility region, no other achievable point
          can dominate the hyperplane $\mc{H}(\ul{\tilde{u}})$, which shows that $\ul{\tilde{u}}$ is PO.\\
Now assume that $\mc{U}_{\mc{G}}^{C}$ is convex. Denote by $u_i^{\mathrm{max}}$ the maximal utility for player
 $i$. Define a set of $K$ points $(\hat{u}_i)_{i \in K}$, such that:
  $\hat{u}_i=(0,\ldots,u_i^{\mathrm{max}} +1,\ldots,0)$. Let $\PP$ be a polyhedron defined as the subset
   of $\R^K_+$ that dominates every hyperplane passing through the OP $\tilde{\ul{u}}$ and a combination of $K-1$ points $\hat{u}_i$. As
$\mc{U}_{\mc{G}}^{C}$ is convex, this polyhedron intersects the
achievable utility region
 in a unique point which is the OP $\tilde{\ul{u}}$. Since the positive orthan at
    the OP is strictly included in the polyhedron,
     the OP is therefore PO.


\section{Proof of Proposition \ref{prop:SE-vs-ptilde}}
\label{sec:app-proof-SE-vs-ptilde}

\emph{Statement (i).} From Sec. \ref{sec:review-stackelberg} we have
that
\begin{eqnarray}
\frac{\tilde{u_i}}{u_i^L} &=  \frac{\frac{f(\tilde{\gamma})}{\tilde{\gamma}}
[1-(K-1)\tilde{\gamma}](1+\beta^*)}{\frac{f(\gamma^*)}{\gamma^*}
[1-(K-1)\gamma^*\beta^*-(K-2)\beta^*]}\\
&\geq  \frac{\frac{f(\tilde{\gamma})}{\tilde{\gamma}}
[\beta^* - (K-1)\gamma^*\beta^* + 1 - (K-1)\gamma^*]}{\frac{f(\gamma^*)}{\gamma^*}(\beta^* - (K-1)\gamma^*\beta^* +
1 - (K-1)\beta^*)}.
\end{eqnarray}
Since by definition $\beta^* \geq\gamma^*$ we readily see that $\frac{\tilde{u_i}}{u_i^L} \geq1$.\\
\emph{Statement (ii).} From Sec. \ref{sec:review-stackelberg} we
have that
\begin{eqnarray}
\frac{\tilde{u_i}}{u_i^F} = \frac{\frac{f(\tilde{\gamma})}{\tilde{\gamma}}
[1-(K-1)\tilde{\gamma}](1+\gamma^*)}{\frac{f(\beta^*)}{\beta^*}[1-(K-1)\gamma^*\beta^*-(K-2)\beta^*]}\\
\geq \frac{\frac{f(\tilde{\gamma})}{\tilde{\gamma}}\left\{1-\tilde{\gamma}
[(K-1)\gamma^*+K-2]\right\}}{\frac{f(\beta^*)}{\beta^*}\left\{1-\beta^*[(K-1)\gamma^*+K-2]\right\}}.
\end{eqnarray}

We want to prove that this ratio is greater than or equal to $1$. We consider
the quantity~:
\begin{footnotesize}
\begin{eqnarray}
\varphi_{[\tilde{\gamma}_K,\beta^*]}(K)&=
\frac{f(\tilde{\gamma}_K)}{\tilde{\gamma}_K}
\left\{1-\tilde{\gamma}_K [(K-1)\gamma^*_K+K-2]\right\}\\
&-\frac{f(\beta^*)}{\beta^*}\left\{1-\beta^*
[(K-1)\gamma^*_K+K-2]\right\}\nonumber\\
&= K(f(\beta^*)-f(\tilde{\gamma}_K))(\gamma_K^*+1)+ \frac{f(\tilde{\gamma}_K)}{\tilde{\gamma}_K}\\
&-\frac{f(\beta^*)}{\beta^*}-(f(\beta^*)-f(\tilde{\gamma}_K))(\gamma^*_K+2)\nonumber\\
&\geq K(f(\beta^*)-f(\tilde{\gamma}_K))(\tilde{\gamma}_K+1)+ \frac{f(\tilde{\gamma}_K)}{\tilde{\gamma}_K}\\
&-\frac{f(\beta^*)}{\beta^*}-(f(\beta^*)-f(\tilde{\gamma}_K))(\beta^*+2)
\end{eqnarray}
\end{footnotesize}
As $\tilde{\gamma}_{K}$ goes to zero as $K$ goes to infinity, ($\tilde{\gamma}_{K}\leq \frac{1}{K-1}$) and by hypothesis $f(0)=0$, $\lim_{x \rightarrow 0} \frac{f(x)}{x}=0$ :
\begin{footnotesize}
\begin{eqnarray}
&\lim_{K \rightarrow
+\infty}\varphi_{[\tilde{\gamma}_K,\beta^*]}(K)\\
&\geq \lim_{K \rightarrow
+\infty}K[(f(\beta^*)-f(\tilde{\gamma}_K))(\tilde{\gamma}_K+1)]\\
&+ \frac{f(\tilde{\gamma}_K)}{\tilde{\gamma}_K}-\frac{f(\beta^*)}{\beta^*}-(f(\beta^*)-f(\tilde{\gamma}_K))(\beta^*+2) \nonumber \\
&\geq \lim_{K \rightarrow
+\infty}K(f(\beta^*))
-\frac{f(\beta^*)}{\beta^*}-(f(\beta^*))(\beta^*+2)=+\infty
\end{eqnarray}
\end{footnotesize}
In conclusion, the sequence
$(\varphi_{[\tilde{\gamma}_K,\beta^*]}(K))_{K \geq 2}$ is strictly
increasing and its limit is $+\infty$. There exists an integer $K_0$ such
that for all $K\geq K_0$,
$\varphi_{[\tilde{\gamma}_K,\beta^*]}(K)>0$. This implies that $u_i(\tilde{\ul{p}}) > u_i^F$.

\section{Proof of Proposition \ref{prop:NE-vs-ptilde}}
\label{sec:app-proof-NE-vs-ptilde}

The goal is to show that $\forall i \in K, \frac{u_i(\tilde{\ul{p}})}{u_i(\ul{p}^*)}
 =\frac{\frac{f(\tilde{\gamma})}{\tilde{\gamma}}
 [1-(K-1)\tilde{\gamma}]}{\frac{f(\gamma^*)}{\gamma^*}[1-(K-1)\gamma^*]}$ is greater than or equal to one. For this, consider the
 function $\phi(x)
=\frac{f(x)}{x}\left[1-(K-1)x\right]$. The derivative of $\phi$ is
$\phi'(x)=\frac{x(1-(K-1)x)f'(x)-f(x)}{x^2}$ which vanishes in
unique point namely, in $\tilde{\gamma}$. Therefore the function
$\phi$ is strictly increasing on $]0,\tilde{\gamma}[$ and strictly
decreasing on $]\tilde{\gamma},+\infty[$ and thus reaches its
maximum in $\tilde{\gamma}$, which concludes the proof.

\section{Proof of Theorems \ref{theo:eq-strategies-fini} and \ref{theo:eq-strategies}}
\label{sec:app-proof-eq-RG}

The proofs are provided in the general case where the PC game is
repeated for different channel realizations (DSPC). At a game stage
$t$ a transmitter has therefore to consider future realizations of
the channels, which are unknown at stage $t$. To tackle this issue
we use a dynamic programming principle \cite{shapley-pnas-1953}, which is standard in repeated game.
Let us define $\bar{u_i}=\max_{\ul{p}}
u_i(\ul{p})=\frac{R_i|g_i|^2f(\beta^*)}{\sigma^2\beta^*}$, the
maximal utility player $i$ can get for a fixed channel gain
$|g_i|^2$. First consider the FRG.
 It is clear that during the last $T_0$ stage, the
 players play the one-shot NE. Thus no
 deviation over this period can be profitable to any player. We
 now consider that a player deviates during the first $T-T_0$ stages.
 His deviation utility is bounded by $\bar{u}_i$ and he will be
 punished at his minmax level in the next stage. Suppose the
 deviator deviates at stage $t\leq T-T_0$. Denote by $\stackrel{\frown}{u_i} = \min_{\ul{p}_{-i}}
 \max_{p_i} u_i(\ul{p})$ with $\ul{p}_{-i} = (p_1,...,p_{i-1},p_{i+1},
 ...,p_K)$. The deviation
 utility of the FRG is upper-bounded as:
 \begin{footnotesize}
 \begin{eqnarray*}
&\sum_{s=1}^{t-1}\tilde{u_i}(\ul{p}(s))+ \bar{u_i}(\ul{p}(t))
 +\sum_{s =t+1}^{ T}\mathbb{E}_{\mb{g}}[\stackrel{\frown}{u}_i(\ul{p}(s))]\\
 &\leq    \sum_{s=1}^{T-T_0-1}\tilde{u_i}(\ul{p}(s))
 +\bar{u_i}(\ul{p}(T-T_0))+\sum_{s =T-T_0+1 }^{T}\mathbb{E}_{\mb{g}}[\stackrel{\frown}{u_i}(\ul{p}(s))]
\end{eqnarray*}
\end{footnotesize}

 The equilibrium condition for the FRG at stage $t$ writes as:
 \begin{footnotesize}
\begin{eqnarray*}
 & \sum_{s=1}^{T-T_0-1}\tilde{u_i}(\ul{p}(s))+\bar{u_i}(\ul{p}(T-T_0))
 +\sum_{s =T-T_0+1 }^{T}\mathbb{E}_{\mb{g}}[\stackrel{\frown}{u_i}(\ul{p}(s))]\\
 &\leq \sum_{s=1}^{T-T_0-1}\tilde{u_i}(\ul{p}(s))+\tilde{u_i}(\ul{p}(T-T_0))
 +\sum_{s =T-T_0+1 }^{T}\mathbb{E}_{\mb{g}}[u_i^*(\ul{p}(s))]
 \end{eqnarray*}
 \begin{eqnarray*}
&\Longleftrightarrow \quad \bar{u_i}(\ul{p}(t)) +\sum_{s =T-T_0+1
}^{T}\mathbb{E}_{\mb{g}}[\stackrel{\frown}{u_i}(\ul{p}(s))]\\
&\leq \tilde{u_i}(\ul{p}(t))+\sum_{s =T-T_0+1
}^{T}\mathbb{E}_{\mb{g}}[u_i^*(\ul{p}(s))]
\end{eqnarray*}
\begin{eqnarray*}
&\Longleftrightarrow  \quad |g_i|^2\frac{f(\beta^*)}{\beta^*}
+\sum_{s
=T-T_0+1
}^{T}\mathbb{E}_{\mb{g}}[\frac{|g_i|^2f(\beta^*)}{\beta^*\left(\sum_{j
\in
\mc{K} \backslash i} p_j^{\max}|g_j|^2+\sigma^2\right)}]\\
&\leq |g_i|^2
\frac{f(\tilde{\gamma})(1-(K-1)\tilde{\gamma})}{\tilde{\gamma}}
+\sum_{s =T-T_0+1
}^{T}\mathbb{E}_{\mb{g}}[|g_i|^2]\frac{f(\beta^*)(1-(K-1)\beta^*)}{\beta^*}.\nonumber
\end{eqnarray*}
\end{footnotesize}

Now we want to show that the last inequality is verified under the
sufficient condition of Theorem $8$. The sufficient condition of
Theorem $8$ implies that:
\begin{footnotesize}
\begin{eqnarray*}
& \eta_i^{\mathrm{max}}\frac{f(\beta^*)}{\beta^*} -
\eta_i^{\mathrm{min}}
\frac{f(\tilde{\gamma})(1-(K-1)\tilde{\gamma})}{\tilde{\gamma}}\\
&\leq
T_0 \left[\;\eta_i^{\mathrm{min}}
\frac{f(\beta^*)(1-(K-1)\beta^*)}{\beta^*}-
 \frac{\eta_i^{\mathrm{max}}f(\beta^*)}{\beta^*\left(\sum_{j
  \in \mc{K} \backslash i} p_j^{\max}\eta_i^{\mathrm{min}}+\sigma^2\right)}\right]
\end{eqnarray*}
\begin{eqnarray*}
&\Longrightarrow
\eta_i^{\mathrm{max}}\frac{f(\beta^*)}{\beta^*}+T_0\frac{\eta_i^{\mathrm{max}}f(\beta^*)}{\beta^*\left(\sum_{j
\in
\mc{K} \backslash i} p_j^{\max}\eta_i^{\mathrm{min}}+\sigma^2\right)}\\
&\leq \eta_i^{\mathrm{min}}
\frac{f(\tilde{\gamma})(1-(K-1)\tilde{\gamma})}{\tilde{\gamma}}
+T_0\eta_i^{\mathrm{min}}\frac{f(\beta^*)(1-(K-1)\beta^*)}{\beta^*}
\nonumber
\end{eqnarray*}
\begin{eqnarray*}
&\Longrightarrow \eta_i^{\mathrm{max}}\frac{f(\beta^*)}{\beta^*}
+\sum_{s =T-T_0 +1}^{T}\frac{\eta_i^{\mathrm{max}}f(\beta^*)}{\beta^*\left(\sum_{j \in
\mc{K} \backslash i} p_j^{\max}\eta_i^{\mathrm{min}}+\sigma^2\right)}\\
 &\leq \eta_i^{\mathrm{min}} \frac{f(\tilde{\gamma})(1-(K-1)\tilde{\gamma})}{\tilde{\gamma}}
 +\sum_{s =T-T_0 +1}^{T}\eta_i^{\mathrm{min}}\frac{f(\beta^*)(1-(K-1)\beta^*)}{\beta^*},\nonumber
 \end{eqnarray*}
 \end{footnotesize}

The worst case scenario for stochastic channel gains implies the
average case scenario. The condition of theorem
\ref{theo:eq-strategies-fini} is sufficient for the desired
equilibrium condition hold at each stage $t$ of the FRG. This
concludes the proof for the FRG. Consider now the equilibrium
condition for the DRG at stage $t$ :
\begin{footnotesize}
\begin{eqnarray*}
&\lambda  \bar{u_i}(\ul{p}^t)+\sum_{s \geq t+1}\lambda(1-\lambda)^{s-t}\mathbb{E}_{\mb{g}}[u_i^*(\ul{p}^s)]\\
&\leq \lambda  \tilde{u_i}(\ul{p}^t)+\sum_{s\geq t+1}\lambda(1-\lambda)^{s-t}\mathbb{E}_{\mb{g}}[\tilde{u_i}(\ul{p}^s)]
\end{eqnarray*}
\begin{eqnarray*}
&\Longleftrightarrow& \lambda
|g_i|^2\frac{f(\beta^*)}{\beta^*}\\
&+&\sum_{s\geq t+1}
\lambda(1-\lambda)^{s-t}\mathbb{E}_{\mb{g}}[|g_i|^2]\frac{f(\beta^*)(1-(K-1)\beta^*)}{\beta^*}\\
 &\leq& \lambda |g_i|^2 \frac{f(\tilde{\gamma})
 (1-(K-1)\tilde{\gamma})}{\tilde{\gamma}}\\
 &+&\sum_{s\geq t+1}
 \lambda(1-\lambda)^{s-t}\mathbb{E}_{\mb{g}}[|g_i|^2]
 \frac{f(\tilde{\gamma})(1-(K-1)\tilde{\gamma})}{\tilde{\gamma}}
 \end{eqnarray*}
\begin{eqnarray*}
&\Longleftrightarrow& \lambda
|g_i|^2\left[\frac{f(\beta^*)}{\beta^*}
-\frac{f(\tilde{\gamma})(1-(K-1)\tilde{\gamma})}{\tilde{\gamma}}\right]\\
&\leq& \sum_{s\geq t+1}\lambda(1-\lambda)^{s-t}\mathbb{E}_{\mb{g}}
\left[|g_i|^2\right]\\
&\times& \left[\frac{f(\tilde{\gamma})(1-(K-1)\tilde{\gamma})}{\tilde{\gamma}}
-\frac{f(\beta^*)(1-(K-1)\beta^*)}{\beta^*}\right]
\end{eqnarray*}
\end{footnotesize}

The equilibrium condition for the DRG can be obtained by following the same reasoning as for the FRG.

\bibliography{biblio}

\vspace{0.5cm}

\begin{footnotesize}
\textbf{Mael Le Treust}
earned his Dipl\^{o}me d'\'{E}tude Approfondies (M.Sc.) degree in Optimization, Game Theory \& Economics (OJME) from the Universit\'{e} de Paris VI (UPMC), France in 2008. He is currently pursuing his Ph.D. degree at the Laboratoire des signaux et syst\`{e}mes (joint
laboratory of CNRS, Sup\'{e}lec, Universit\'{e} de Paris XI), Gif-sur-Yvette, France. He was also a Math TA at the Universit\'{e}
de Paris I (Panth\'{e}on-Sorbonne) and Universit\'{e} de Paris VI (UPMC), France. His research interests include game theory, wireless communications and information theory.

\textbf{Samson Lasaulce}
received his BSc and Agr\'{e}gation degree in Applied Physics from \'{E}cole Normale Sup\'{e}rieure (Cachan) and his MSc and PhD in Signal Processing from \'{E}cole Nationale Sup\'{e}rieure des T\'{e}l\'{e}communications de Paris (ENST). He has been working with Motorola Labs (1999, 2000, 2001) and France T\'el\'ecom R+D (2002, 2003). Since 2004, he has joined the CNRS and Sup\'{e}lec and is Charg\'{e} d'Enseignement at \'{E}cole Polytechnique. His broad interests lie in the areas of communications, signal processing, information theory and game theory for wireless communications.
\end{footnotesize}

\end{document}